\documentstyle[aps,prb,floats,epsf]{revtex} 
\begin{document}
\draft
\flushbottom

\twocolumn[\hsize\textwidth\columnwidth\hsize\csname
@twocolumnfalse\endcsname

\title{Ab-initio structural, elastic, and vibrational 
properties of carbon nanotubes}

\author{Daniel S\'anchez-Portal, Emilio Artacho, and Jos\'e M. Soler}
\address{Departamento de F\'{\i}sica de la Materia Condensada and 
Instituto Nicol\'as Cabrera, C-III, \\
Universidad Aut\'onoma de Madrid, 28049 Madrid, Spain}
\author{Angel Rubio}
\address{Departamento de F\'{\i}sica Te\'orica, 
Universidad de Valladolid,
47011 Valladolid, Spain} 
\author{Pablo Ordej\'on}
\address{Departamento F\'{\i}sica, Universidad de Oviedo, 33007 Oviedo,
Spain}

\date{\today}

\maketitle

\begin{abstract}
A study based on {\it ab~initio} calculations is presented on the 
structural, elastic, and vibrational properties of single-wall carbon
nanotubes with different radii and chiralities. These properties are
obtained using an implementation of 
pseudopotential-density-functional theory 
which allows calculations on systems 
with a large number of atoms per cell. 
Different quantities are monitored versus tube radius.
The validity of 
expectations based on graphite is explored down to small radii, where 
some deviations appear 
related to the curvature-induced rehibridization of the 
carbon orbitals. 
Young moduli are found 
to be very similar to graphite and do not exhibit a systematic 
variation with either the radius or the chirality.
The Poisson ratio also retains
graphitic values except for a possible slight reduction for small radii.
It shows, however, chirality dependence.
The behavior of characteristic phonon branches as the breathing mode, 
twistons, and high-frequency optic modes, is also studied, the latter 
displaying a small chirality dependence at the top of the band. The results 
are compared with the predictions of the simple zone-folding
approximation.
Except for the known defficiencies of the
zone-folding procedure in the low-frequency vibrational regions, it
offers quite accurate results, even for relatively small radii.
\end{abstract}

%\pacs{PACS numbers: 36.40.+d; 81.20.Sh; 63.20Dj (CHECK) }
]

\narrowtext

\section {INTRODUCTION}
Carbon nanotubes have excited a considerable interest in the condensed-matter
and materials research communities in the last few years and much experimental
and theoretical work has been devoted to them as prototype of one-dimensional
ordered systems with promising technological applications~\cite{Carbon}. 
Electronic transport in conducting nanotubes is one of the issues that has 
attracted more attention, especially after the developments that made 
possible the sinthesis of large quantities of single wall nanotubes (SWNT)
forming crytalline ropes~\cite{Smalley,Journet}.
Experiments showed a peculiar metallic behaviour 
above 35~K which was understood in terms of
the coupling between the conduction electrons and long wavelength twistons,
i.e., torsional shape vibrations~\cite{twistons}.
The quantitative understanding and characterization of this
and other related phenomena require the detailed 
knowledge of 
both the structure and 
the vibrations of these tubes. 

The structure of carbon nanotubes is qualitatively well known through
the simple construction of rolling a perfect graphene sheet, where only
one parameter is to be determined: the lattice parameter or a bond-length. 
The symmetry of the tubes is less restrictive than in graphite 
and several 
parameters are needed to determine completely the structure. 
Among other things, these parameters define the differences between 
inequivalent bonds, which condition the position of the Fermi surface 
in the conducting armchair tubes. It is very difficult to obtain direct 
experimental information for the structure, and very little 
theoretical information has 
been given so far~\cite{JYYi,Xavier,Kresse_breathing}. 

Besides possible nanotechnological applications, carbon nanotubes are
promising candidates for composite materials where their low weight and
very high Young modulus can be of use. Their elastic properties have thus
received considerable attention as well. 
The nanotube's unusual strength arises from a 
combination of high stiffness and 
a extraordinary flexibility and resistance 
to fracture~\cite{Iijima,Yakobson,Falvo}. 
The Young modulus of SWNT has been measured from the amplitude 
of their thermal vibrations~\cite{Treacy} and by measuring the bending 
force of a pinned nanotube by an atomic force microscope~\cite{Wong}.
The experimental results
give a Young modulus in the range of one TPa, similar to the one of
graphite when pulled parallel to the sheets, but the experimental 
uncertainity is quite high. There is also a dispersion of theoretical 
values~\cite{Iijima,Yakobson,Mintmire,Ruoff,JPLu,Edu}
in the literature
for this quantity corresponding to different 
approximations but also to different definitions of the effective
sectional area. 
 
Raman experiments~\cite{Rao} offer valuable
information for the vibrations of ropes of single-wall armchair 
($n,n$) tubes. The use of resonant Raman 
scattering~\cite{Raman_resonante} allows to 
discriminate the vibrations stamming from tubes of different diameters
by looking mainly at the A$_{1g}$ breathing mode that exhibit a strong 
dependence on tube-diameter. Furthermore, the optical E$_{2g}$ phonon
peak shows size-dependent multiple splittings that are nearly independent of
chirality and can be used to determine the tube-diameter~\cite{Kasuya}.
Raman scattering, however, 
is limited to the neighborhood of the center of the Brillouin zone, and 
additional selection rules limit the number of vibrations detectable to
seven per tube type. The theoretical characterization of the vibrational modes
can complement this knowledge allowing to correlate tube diameter to
a specific vibrational frequency.
There is theoretical information available
based on empirical force constants~\cite{Rao} and tight-binding
Hamiltonians~\cite{Yu,Menon}, but a first-principles theoretical reference
is still lacking~\cite{celtiberia}
(except for the breathing mode~\cite{Kresse_breathing}).

The strong similarity of the chemistry of carbon nanotubes to graphite
allows theoretical analyses to be done based on empirical methodologies 
imported from studies on graphite. They range from the direct 
zone-folding~\cite{zf} of the results for graphite to the quantum-mechanical 
studies based on tight-binding Hamiltonians fitted to graphite 
properties~\cite{Yu,Menon,Ho}. Effective interatomic potentials~\cite{Mintmire}, 
force-constant models~\cite{JPLu} or non-orthogonal tight-binding~\cite{Edu}
have also been tried. The performance
of the different techniques varies, from the qualitative picture offered
by zone folding, with intrinsic defficiencies for low frequencies, to the
very quantitative results of tight-binding approaches. The curvature of
the tubes, however, disturbs the chemistry in a way that can cause 
the deviation, from the graphite based description, for narrow tubes. 
Zone folding and 
force-constants neglect curvature all together. Model potentials
can only account for the different distances among the atoms. 
Tight-binding captures part of the chemical strain through the 
geometry-dependence of its electronic matrix elements, even though
their absolute value depend on the electronic structure of graphite. 
It is then important to be able to compute the different properties 
for any tube radius using a tool that does not depend 
on a fit to graphitic properties, 
so as to study the narrow-tube properties
with the same degree of accuracy as the wide ones.

In the present work we present an {\it ab~initio} study of the structural,
elastic, and vibrational properties of single-wall carbon nanotubes
for different diameters and chiralities to address the points mentioned
above. The behavior of the different properties is monitored versus
tube radius, and the performance of the different approximations
used in the literature is studied to ascertain on their different
ranges of applicability. Section II describes the methodological
details. The results and discussion are presented in Section III,
to finish with the concluding remarks in Section IV.

\section {COMPUTATIONAL SCHEME}

The first-principles scheme used in this study was described in detail
elsewhere~\cite{ordenN}, where it was already applied and tested on
large fullerene molecules. It is based on 
the Local Density Approximation (LDA) to Density Functional Theory~\cite{LDA}.
Core electrons are replaced by non-local, norm-conserving 
pseudopotentials~\cite{pseudo}, whereas valence electrons are described
within the Linear Combination of Atomic Orbitals (LCAO)
approximation. 
In this work we have used a minimal basis set of one $s$
and three $p$ orbitals per carbon atom. Although this basis
is certainly not complete, it provides a sufficiently accurate
description of the systems and effects that we intend
to study.

The radial shape of the atomic basis functions is chosen
according to the prescription of Sankey and Niklewski~\cite{PAO}.
This consists in using the solution of the atom with the 
pseudopotential as a basis for the LCAO calculation;
these pseudoatomic orbitals (PAO) are calculated with the
boundary condition that they vanish outside a given radius
$r_c$. The PAO's are therefore slightly excited, since their
kinetic  energy is increased due to the vanishing boundary condition.
They are strictly localized, which is very
advantageous from the computational point of view.
For a review of the quality of these PAO bases for 
electronic structure calculations, we refer the
reader to Ref.~\onlinecite{jpc-bases}.
In this work, we have used a radial cutoff of $r_c=4.1$ Bohr.

The use of atomic orbitals,
and a combination of efficient techniques,
allow us to calculate the LDA Hamiltonian with an order-$N$
effort (i.e., a cost that scales linearly with the number
of atoms, both in time and in memory)~\cite{ordenN}. 
This makes possible to reach system sizes with a much larger
number of atoms than the standard techniques. 
Some of the Hamiltonian matrix elements are computed by
interpolation of pre-calculated two and three center integral
tables, whereas others are obtained by
direct integration in a real space grid. 
The fineness of the grid is expressed by
the maximum kinetic energy of a plane wave
that can be represented in it, as is usually
done in plane wave calculations. Note that, in our case, the
grid is used to represent the charge density, and not the
wavefunctions. In this work we have used a cutoff of
60 Ry for the grid integrations.

The solution of the Hamiltonian matrix can
be done by straight diagonalization (which is an
${\cal O}(N^3)$ operation), or using recently developed order-$N$
techniques~\cite{ordejon1,ordejon2}, 
which are very advantageous for systems with
large numbers of atoms (tipically, the crossover between the
two solutions is around 100 atoms, for calculations with minimal
bases).
The systems under study in this work are in the range between
80 and 200 atoms, and therefore diagonalization of the Hamiltonian
is feasible and competitive with an order-$N$ technique.
We have used diagonalization throughout this work.

Calculations were performed for the following tubes: (4,4), (6,6), (8,8),
(10,10), (10,0), and (8,4). They were all considered as isolated, infinitely
long tubes. For that purpose, we
used periodic boundary conditions on a 
supercell geometry with sufficient
lateral separation among neighboring tubes. 
For the purpose of sampling the Brillouin zone 
in the direction of the tube axis, as well as for the computation
of the force constant matrix for phonon calculations (see below),
we used supercells containing several unit cells in the axis direction.
The supercells consisted of five unit cells 
for the armchair ($n,n$) tubes (80, 120, 
160, and 200 atoms for the (4,4), (6,6), (8,8), and (10,10) tubes, 
respectively), three unit cells for the zig-zag (10,0) tube (120 atoms), 
and one unit cell for the chiral (8,4) tube (112 atoms). The supercell
length in the tube-axis direction is similar for all of them, the number
of atoms changing because of the different tube radii.
Only 
the $\Gamma$ point of the supercell was used in the 
phonon calculations, although more complete
$k$-point samplings were tested showing no significant differences with
the $\Gamma$ point results.

The atomic structures of the tubes were obtained by careful minimization
of the total energy by means of the calculated Hellman-Feynman forces,
including  Pulay-like corrections to account for the fact that the 
basis set is not complete and moves with the atoms~\cite{ordenN}.
The minimizations were performed using a dissipative
molecular-dynamics algorithm which allows the geometry optimization
with no symmetry conditions imposed. The residual forces in the
optimizations were always smaller than 0.04~eV/\AA.
The energy was minimized and the structure relaxed for different
values of the lattice constant along the tube axis. This procedure 
allows us to determine the most stable lattice constant
along the tube, as well as the
the Young modulus and the Poisson ratio. 

For the phonon calculations, we first obtain the
force constant matrix in real space using a 
finite difference approach~\cite{Pablo,PRL_alemanes}. 
We used finite atomic displacements of 0.02~\AA,
and the force constants were taken as the average
of the results obtained with positive and negative
displacements, to elliminate anharmonic effects. 
All atoms are equivalent by symmetry, in the nanotubes,
and therefore we only calculated the force constants
for one of the atoms in the supercell, and generated 
the rest of the matrix using the symmetry operations.

The force-constant matrix has to be computed
between a given atom and all the rest in the system.
However, it is known that the force-constants
decrease
rapidly with distance (in non-polar systems), so that
only the elements of atoms sufficiently close need to 
be computed. To do so, we set up a supercell large enough that
a sufficient number of neighbors in the tube axis
direction is included.  It must be also kept in mind
that, in the supercell geometry, a given atomic displacement
in the central cell is always accompained by the same
displacement of all the images. The supercell must therefore
be large enough so that the effect of the image displacement
is neglegible. The cells discussed above were built
with these observations in mind.
Once the force constant matrix in real space has been obtained,
we calculate the dynamical matrix in reciprocal
space, and diagonalize it to obtain the phonon
modes and frequencies as a function of the 1-D crystalline
momentum vector in the direction of the tube.

\section{RESULTS AND DISCUSSION}

\subsection {Structural properties}

\begin{figure}[!]
\narrowtext
\centering
\epsfxsize=0.95\linewidth
\epsffile{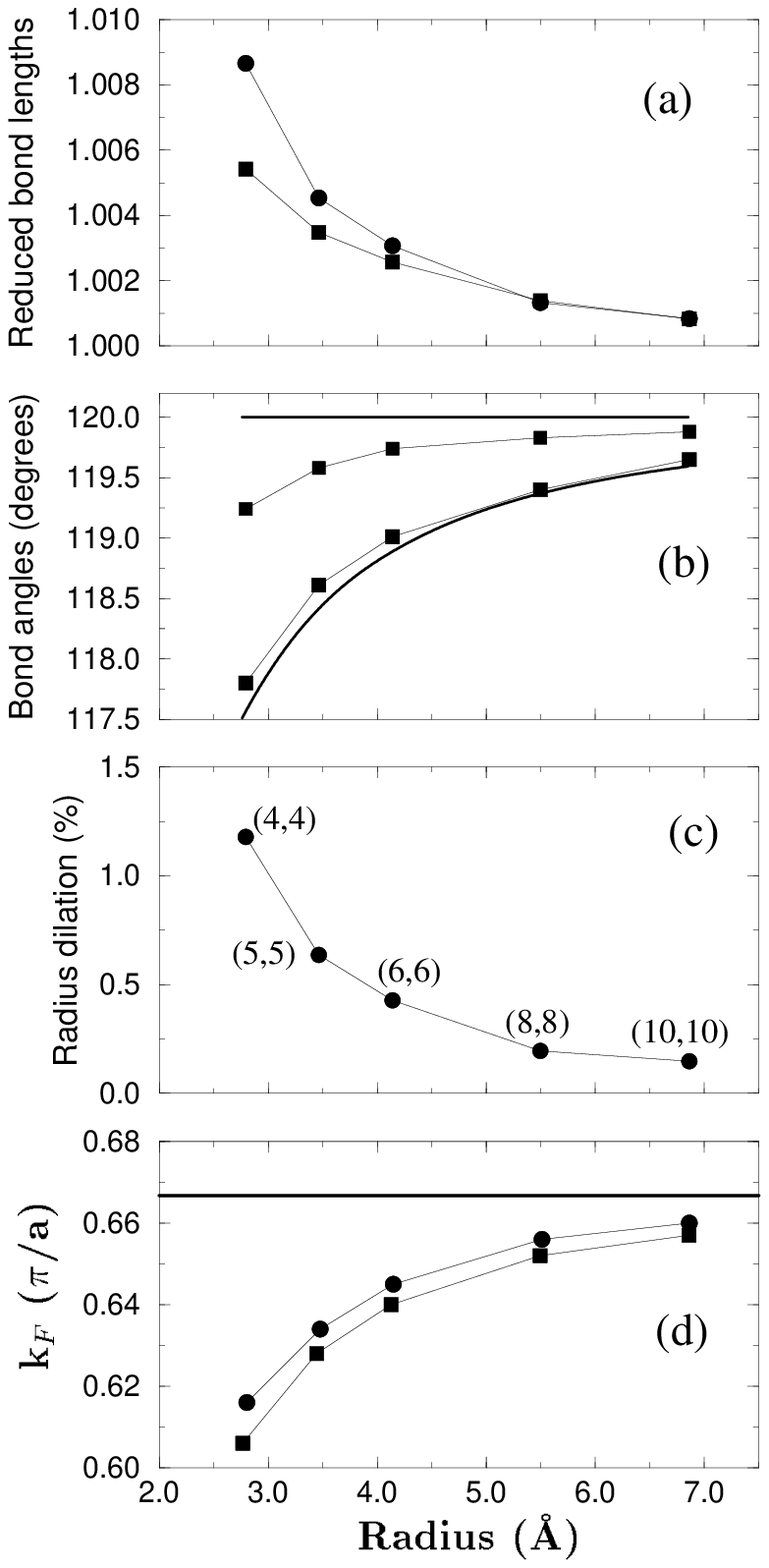}
\vspace{0.5 cm} 
\caption[]{Structure variations with tube radius. (a) Lengths
of the two inequivalent bonds 
in ($n,n$) tubes in units of the bond length
in graphite. Circles stand for the bond 
perpendicular to the tube axis, and
the squares for the other inequivalent bond-length.
(b) The two inequivalent bond angles in $(n,n)$
tubes. Continuous lines show the angles resulting from
an ideal graphene rolling. (c) Tube radius 
dilation as compared to the ideal
radius. (d) Position of the Fermi surface ($k_F$) in the Brillouin zone.
Squares and circles stand for the ideal
and relaxed structures, respectively.
The continuous line indicates the ideal zone-folded value.}
\label{distkf}
\end{figure}

We first study the equilibrium structural properties
of the nanotubes. As a reference, we have first computed the
equilibrium bond distance for a single graphene plane, for which
we obtain a value of $d$=1.436 \AA, close to the
experimental~\cite{Baskin} (1.419 \AA)
and LDA plane-wave calculations~\cite{Schabel} (1.415 \AA) in graphite.
Our value is slightly larger than the one obtained with plane-waves, 
the small difference being due to the basis set.
For all the tubes studied here we have found that the average
carbon bond-length is within 1\% of the graphitic value.
No appreciable symmetry-breaking distortions have been observed.
However, by symmetry,
for the $(n,n)$ and $(n,0)$ tubes there are two inequivalent bonds, 
three in general, for chiral tubes.  
Fig.~\ref{distkf} (a) shows the value of these two bond lengths as a function
of tube radius (in units of the graphene bond length) for
the $(n,n)$ tubes studied here. 
The differences between these bonds and graphite, and between 
the two kinds of bonds, are small but significant. Two effects are
apparent: $(i)$ both bond-lengths 
increase (as compared with the graphene reference) 
with decreasing tube radius; $(ii)$
the difference between them also increases with decreasing radius.
Both effects can be easily understood in terms of the 
rehibridization 
and the weakening of the $\pi$ bonds 
induced by the curvature~\cite{Xavier}.
As shown in Fig.~\ref{distkf} (a), the longer
bond is the one perpendicular to the tube axis in the (n,n) tubes. 
This is in contrast with 
the results of Ref.~\onlinecite{Kresse_breathing}. This discrepancy 
is due to the small k-point sampling used in that work~\cite{distancias}.

Fig.~\ref{distkf} (b) shows the variation of the bond angles with
the tube radius for the $(n,n)$ tubes. 
As for the bond distances, there are two
inequivalent bond angles in $(n,n)$ tubes.
The behavior of the bond angles is very similar to
the one expected from ideal 'rolling' of a graphene
plane to form the tube, also shown 
in Fig.~\ref{distkf} (b). In this case, one of the
angles would maintain a value of 120$^\circ$, whereas
the other would decrease for smaller tube radii,
leading to the increasing tube curvature. In the fully relaxed
structures, both angles are smaller than 120$^\circ$,
so that the curvature stress is more distributed around the
tube area. Since the ideal 120$^\circ$ angle now
shares some of the cost of the curvature, the other angle
increases slightly from its ideal value.
Nevertheless, we see that deviation from the ideal behavior
is only relevant for the tubes with small radius, being
almost negligible for tubes larger than $(6,6)$.

As a consequence of the increase in bonding distances
shown in Fig.~\ref{distkf} (a), the actual tube
radii are slightly larger than those resulting from an ideal
rolling of the graphene sheet. This is 
plotted in Fig.~\ref{distkf} (c).
Again, this deviation is more pronounced for the tubes with smaller
radius, and the behavior approaches the ideal for large radii tubes.

The structural distorsions 
affect the electronic 
structure of these metallic tubes.
Although the study of the electronic properties is not
the focus of this work, it is worth discussing what is
the effect of the structural parameters on the Fermi level, 
since this has implications for the physics
of these systems, for example, in the interpretation of very recent 
scanning tunneling spectroscopy experiments 
on short SWNT~\cite{Venema}.
In a zone-folding model based on the graphene band structure,
the Fermi level of all $(n,n)$ tubes would be located
at a wave vector of ${2 \over 3} {\pi \over L}$ (with $L$ being the
lattice constant in the tube axis). For the real tube 
structure, the position of the Fermi level will change due
to the reduced symmetry, with its two inequivalent bonds
(those perpendicular to the tube axis, and those non-perpendicular).
In a simple H\"uckel model this reduced symmetry is described
with two different 
hopping interactions $t_\perp$ and $t'$, both different
from the ideal graphene hopping $t$. The Fermi wave vector
in this case is $k_F = {2 \over L} \arccos ({t_\perp \over 2 t'})$.
In real tubes, the Fermi surface moves because of two reasons:
$(i)$ The rolling of the graphene plane to
form the tube originates a change of the electronic potential
in the inequivalent bonds of the tube,
even in the case where the structure is taken
as the ideal graphene rolling, i.e. with both bonds kept
at the same graphene bond length. 
$(ii)$ The difference in bond lengths for the two inequivalent
bonds in the relaxed structure change 
the hopping
matrix elements. 
In order to distinguish these two effects, we show
in Fig.~\ref{distkf} (d) the calculated values of
the Fermi wave vector for the fully relaxed structures,
and for the ideal graphene rolling tubes,
obtained with our LDA-LCAO formulation. 
We see that the main effect of the Fermi level
shift is the rolling to form the tube, whereas
the difference in bond lengths brought by the
structure relaxation gives only a minor correction
tending to bring $k_F$ closer to the
graphene value. We see that the deviation from the
graphene Fermi level are relatively large, 
increasing with decreasing tube radii.

\subsection{Elastic properties}

\subsubsection{Strain energy}

Fig.~\ref{fig2} (a) shows the strain energy per atom
(energy relative to a planar graphene sheet) as a function
of the radius of the tube. The data follows
quite well the behavior expected from classical
elasticity theory~\cite{Tibbetts}, $ E_{st}=C/r^2 $, where $r$ is the radius
of the tube and $C$ is a constant that depends on the Young modulus $Y$ and 
thickness $h$ of the wall in a model tube: $C=Yh^3a/24$. A least squares
fit to the results of the $(n,n)$ 
tubes yields a value of $C=2.00$ eV\AA$^2$/atom, in very 
good agreement with 
recent LDA calculations~\cite{Kresse_breathing}. 
For the other two tubes studied, (8,4) and (10,0), 
we obtain slightly larger values (2.15 and 2.16 eV\AA$^2$/atom,
respectively). 

It is to some extent suprising that the predictions
from elasticity theory are so closely followed
by the detailed {\em ab~initio} calculations.
In fact, the assignment of a thickness $h$ for
a single atomic layer is not a well defined procedure.
Adams {\em et al.}~\cite{Adams} provided
an alternative explanation based on microscopic arguments.
They use a very simplified model in which the energetics
of many different fullerene structures depend on a single structural
parameter: the {\em planarity} $\phi_\pi$, which is the angle
formed by the $\pi$ orbitals of neighbour atoms. 
Asuming that the change in total energy is mainly
due to the change in the $t_\pi$ interaction between
these orbitals, and that this change is proportional
to $\cos{\phi_\pi}$, the $r^{-2}$ behaviour is predicted.
Adams {\em et al.} obtained a value of $C=2.12$ eV\AA$^2$/atom
using non-selfconsistent first-principles calculations. 
Previous calculations using Tersoff and 
Tersoff-Brenner potentials~\cite{Mintmire} predict the same dependence and 
gave a value of $C\approx1.5$ eV\AA$^2$/atom and  $C\approx1.2$ eV\AA$^2$/atom,
respectively.

We note in Fig.~\ref{fig2} (a) that the
$(n,n)$ tubes are energetically more stable
as compared to other chiralities with the same radius.
This difference is, however, very small and
will decrease as the tube diameter increases due to
the $r^{-2}$ behaviour.
This is expected, since in the limit of large radii
the same graphene limit is obtained, regardless of
chirality.

\subsubsection{Young modulus} 
\label{young_section}

\begin{figure}[!]
\narrowtext
\centering
\epsfxsize=0.95\linewidth
\epsffile{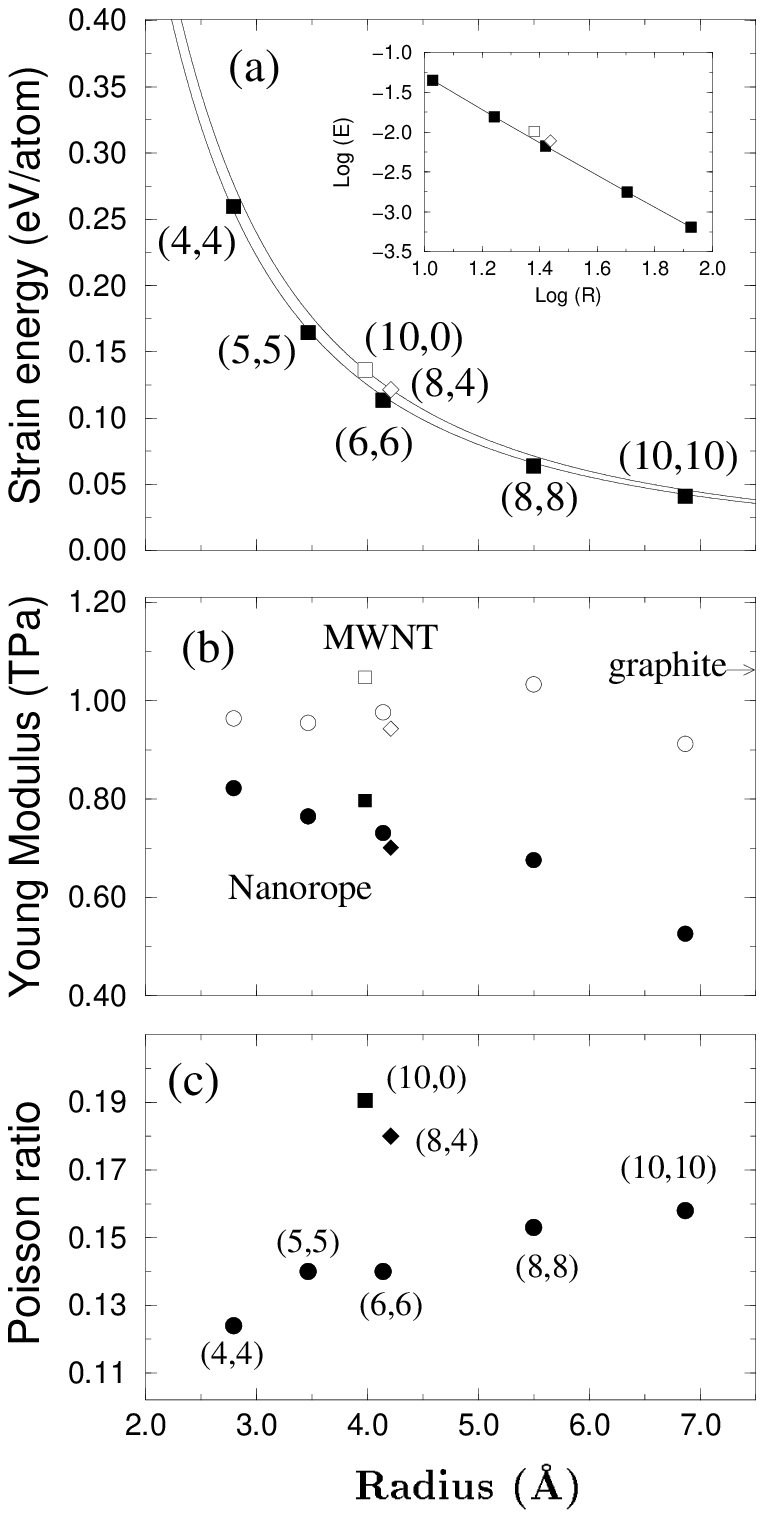}
\vspace{0.5 cm}
\caption[]{(a) Strain energy versus tube radius.
Solid line drawn across the ($n,n$) data corresponds
to a least square fit to the $C/r^2$ behaviour.
The two $C/r^2$ functions passing
through the (8,4) and (10,0) data are also shown
(in this scale they appear as one line).
The $r^{-\alpha}$ behaviour is
clearly shown in the inset.
The value obtained for $\alpha$ from the
logarithmic fit is 2.05 $\pm$ 0.02.
(b) Young modulus versus tube radius.
Open symbols for the {\it multi-wall} geometry,
and solid symbols for the
single-wall-nanotube crystalline-rope configuration.
The experimental value of the c$_{11}$ elastic constant
of graphite is also shown.
(c) Poisson ratio versus tube radius.}
\label{fig2}
\end{figure}
Measurements of the amplitude of the intrinsic thermal vibrations of
the tubes~\cite{Treacy} allow an undirect determination of the 
Young modulus, yielding  
an average value of 1.8 TPa. There
is a large uncertainty in the results, however, with values ranging from 
0.40 TPa to 4.15 TPa. Furthermore, measurements of 
the restoring force on bent nanotubes with an atomic force microscope
give an average value of 1.28$\pm$0.59 TPa.  Since the 
corresponding elastic constant for graphite ($c_{11}$)
is also of the order of 1 TPa,
it is still a question whether the
rolling of the graphene plane to obtain nanotubes 
increases or decreases its stiffness.

Theoretical calculations can help in solving 
the issue, but the calculations performed so far
also show a great dispersion in their predictions. 
Some theoretical estimates report values of the order of 5 TPa~\cite{Tomanek}
based on an empirical Keating force-constant model for finite-capped 
$(5,5)$ tube. This unreasonably high value can be due to the
small size of the aggregates used to describe the
tube, with only up to 400 atoms.
Yakobson {\em et al.}~\cite{Yakobson}, in a study of the structural
instabilities of SWNT for large deformations, and using 
Tersoff-Brenner potentials, obtain 
an estimate for the 
Young modulus of about 5.5 TPa,
by fitting their results
to the continuum elasticity theory. 
However, 
D. H. Robertson et al.~\cite{Mintmire}, using
the same potential, report a small weakening of 
the stiffness of nanotubes as the diameter decreases,
and therefore a Young modulus for the nanotubes  smaller
than the one of graphene. They find a systematic dependence
with the chirality which, although being small,
increases with decreasing tube radius.
J. P. Lu~\cite{JPLu}, using a force constant model 
fitted to reproduce the phonons and elastic constants
of graphite, obtains elastic properties which are essentially 
independent
of helicity and tube radius, and comparable to
those of graphene (with values of the Young modulus
below 1 TPa).
Finally, recent tight-binding~\cite{Edu}
calculations give values of the order of 1 TPa for the SWNT,
quite insensitive to the chirality of the nanotube, and 
mainly determined by the tubule diameter, approaching the
graphitic limit for diameters of $\sim$ 1.2~nm.

Part of the discrepancies in the theoretical
results discused above is merely due to a different
definition of the Young modulus in these systems.
From the point of view of elasticity theory, the
definition of the Young modulus involves the specification
of the value of the {\em thickness} of the tube wall.
As discussed in the previous Section, it is not
clear how to define this width for a SWNT, where
the wall is composed of only one shell of atoms.
The anomalously large value obtained by Yakobson {\em et al.}~\cite{Yakobson}
is due to an assignment of a value of $h=0.6$ \AA~for
the thickness of the graphene plane, which is obviously too 
small. Other authors~\cite{JPLu,Edu} have used the graphite interlayer
spacing of 3.4 \AA~in their calculations.

In order to avoid this definition problem, we have analized
our results on the elastic stiffness of the nanotubes
using the second derivative of the strain energy
with respect to the axial strain: $d^2 E\over{d {\epsilon}^2 }$.
To obtain this quantity for the different tubes, we have
performed structural relaxations for the nanotubes, 
subject to deformations between -1.0\% and 1.0\%,
at intervals of 0.13\%.
The total energies for deformations in the interval
(-0.75\%, 0.75\%) where fitted to a third order polynomial,
and $d^2 E\over{d {\epsilon}^2 }$ was obtained from the
second derivative at zero strain.
The results are shown in Table~\ref{table:d2e}. We have estimated
the numerical error of these results by taking different
intervals in the fitting procedure: (-0.5\%,0.5\%) and (-1.0\%,1.0\%),
and obtain that the typical uncertainty of the calculation 
of $d^2 E\over{d {\epsilon}^2 }$ is of the order of 10\%.
From Table~\ref{table:d2e} we see that the average value 
in the tubes is about 56 eV. The variation between tubes
with different radii and chirality is very small, and always
within the limit of accuracy of the calculation. We therefore
can conclude that the effect of curvature and chirality
on the elastic properties is small.

\begin{table} [!t]
\caption[ ]{Calculated values of $d^2 E / d^2 \epsilon$
for different tubes. The values for a graphene plane obtained
with different supercells are also shown (see text
for details).}

\begin{center} 
\begin{tabular}{lcc} 
Tube    &  R(\AA) &  $d^2 E / d^2 \epsilon$ (eV) \\
\hline 
(4,4)   &  2.794  &  56 \\
(5,5)   &  3.463  &  55 \\
(6,6)   &  4.140  &  56 \\
(8,8)   &  5.498  &  59 \\
(10,10) &  6.864  &  52 \\
\hline
(8,4)   &  4.211  &  54 \\
\hline
(10,0)  &  3.979  &  60 \\
\hline
Graphene (4,4)   &  --  &  50 \\
Graphene (10,10) &  --  &  54 \\
Graphene (rect.) &  --  &  60 \\
\end{tabular}
\end{center}
\label{table:d2e}
\end{table}

For comparison, we have also calculated the 
values of  $d^2 E\over{d {\epsilon}^2 }$ for a
single graphene sheet. Three calculations were
done, with different supercells.  
In all
the cases, only the $\Gamma$ point of the supercells
was used reproducing the $k$-point sampling of the
tube calculations.
Two of them were
the unrolled equivalents of the supercells used for
the (4,4) and (10,10) tubes. These cells were used 
to allow for a direct check of 
the curvature effects, comparing the planar and tubular
geometry.
The third one was a rectangular (nearly square) cell 
of 17.41 \AA $\times$ 17.23 \AA, with 112 atoms, thus
providing a more uniform $k$-point sampling, and, therefore, 
a more confident estimation of the elastic constant.
The results
(also displayed in Table~\ref{table:d2e}) 
clearly show that there are no appreciable differences
between the results obtained for the nanotubes and
those of graphene, the differences being within
the uncertainty of the calculation.
These data confirm that the effect of curvature on the 
Young modulus of the SWNT 
is small down to radii of 
the order of, at least, 2.8~\AA.
 
Our results are in good agreement with
those obtained by Robertson et al.~\cite{Mintmire} 
using Tersoff-Brenner potentials, who find values
around 59 eV/atom, with very little dependence 
with radius and/or chirality. 
Futhermore, we can obtain an experimental
estimate of this quantity using the elastic
constant~\cite{Mintmire} $c_{11}=1.06$~TPa  of bulk graphite,
from which we obtain 
${d^2 E\over{d {\epsilon}^2 }} \simeq c_{11}V_a = 58.2$ eV/atom 
(where $V_a$ is the atomic volumen in graphite).
This value agrees well with the results obtained here,
and with those of Robertson {\em et al.}~\cite{Mintmire}

If one insist in calculating the Young 
modulus using the standard bulk definition, instead
of the well defined second derivative used above,
one must choose a definition of the effective area per 
carbon atom.
Here we calculate the Young moduli considering two
different geometries: $(i)$ a {\it multiwall} like geometry,
in which the normal area is calculated using the
wall-wall distance as the one in multiwall tubes,
which is very approximately equal to the one of graphite,
i.e. the effective {\it thickness} of the the tube wall 
is taken to be 3.4~\AA,  
and $(ii)$ a rope configuration of single wall tubes,
where the tubes would be arranged forming
an hexagonal closed packed lattice, with a lattice constant
of (2$r$ $+$ 3.4~\AA), being $r$ the tube radius.
The results are shown in Fig.~\ref{fig2} (b).
For the crystalline rope geometry, the decrease of 
the Young modulus with increasing the tube radius 
is due to the quadratric increase of the effective area 
in this configuration,
while the
number of atoms increases only linearly with the tube diameter.
The computed values for the SWNT ropes are, however,
still very high, 
even comparing with other carbon fibers.~\cite{carbon_fibers}

\begin{figure*}[!t]
\centering
\epsfxsize=0.9\linewidth
\epsffile{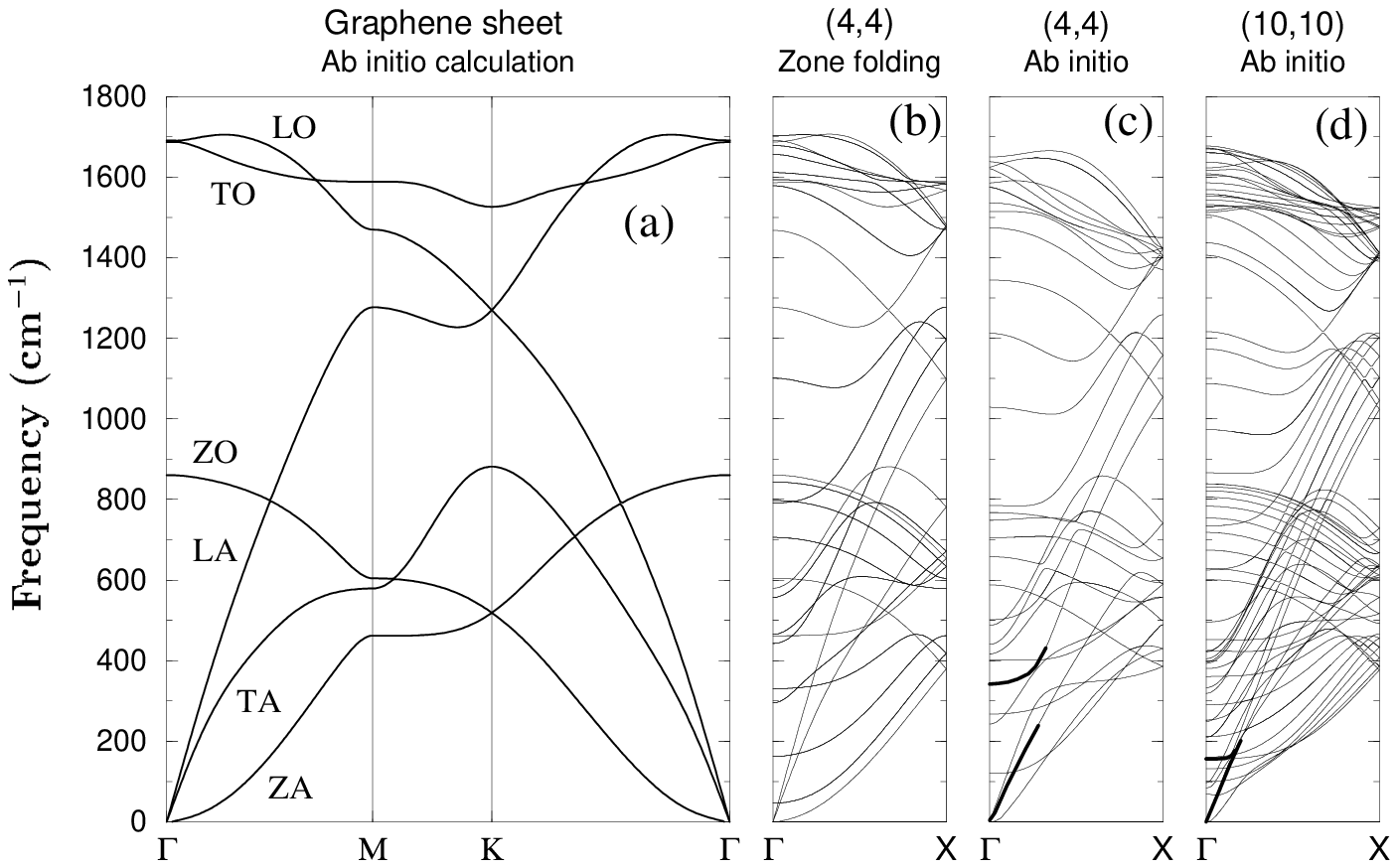}
\vspace{0.5 cm}
\caption[]{(a) Calculated phonon structure for a graphene sheet.
(b) Zone folding result for the (4,4) tube, obtained from the
graphene {\it ab initio} phonons in panel (a).
{\it Ab initio} dispersion relations for the (4,4) (c), and
the (10,10) (d) nanotubes.
In (c) and (d),
thicker lines are used to mark two special branches:
the acoustic band is a twiston mode (torsional shape
vibrations), the other (with finite frequency
at $\Gamma$) is the breathing mode. }
\label{dispersions}
\end{figure*}

\subsubsection{Poisson ratio}

The Poisson ratio $\nu$ is given by the variation of the radius 
of the SWNT resulting from longitudinal deformations
along the tube axis:
\begin{equation}
{\Delta r \over r} = - \nu {\Delta l \over l}
\end{equation}
where $l$ is the tube length. We have calculated $\nu$ for
the tubes under study, and find that in all cases
the Poisson ratio is positive: an elongation of the tube reduces
its diameter. The results are shown in Fig~\ref{fig2} (c).
We obtain values around 
$\nu=0.14$ (from 0.12 to 0.16)
for the armchair $(n,n)$ tubes, and a little larger for
other chiralities: 0.19 for (10,0) and 0.18 for (8,4).
The uncertainty of the obtained values is of the order
of 10\%.
These results reveal a slight decrease of the Poisson ratio
with the tube radius, and a stronger dependence with 
chirality. 
Our results are close to the value of $\nu=0.19$
obtained by Yakobson et al.~\cite{Yakobson} using
Tersoff-Brenner potentials, but
considerably smaller than the value $\nu=0.28$
given by  Lu~\cite{JPLu} with a force constant model
and $\nu=0.26$ from a tight-binding calculation.~\cite{Edu}
The corresponding magnitude along the basal plane in graphite
is $\nu=0.16$.~\cite{Poison_grafito,pw_pois}

\subsection {Vibrations}

In our study of the vibrational properties of SWNT,
we have first computed the phonon spectrum
of a single graphene plane, which is shown
in Fig.~\ref{dispersions} (a). This will serve as a
test of the accuracy of the calculation method, and
as a reference for the interpretation of the 
nanotube results. Also, the graphene phonon
structure is needed to obtain nanotube phonons
within the zone-folding approach.

The calculation of the graphene phonons has been 
performed using the nearly square supercell of 112 atoms 
described in previous sections.
The calculated {\em ab~initio} phonon dispersion 
curves are, in general, in quite good agreement 
with experiments for graphite.~\cite{Oshima}
The most remarkable disagreement
is for the frequency of the higher optical bands,
which at $\Gamma$ is around 1690 cm$^{-1}$ in our calculations,
and 1580 cm$^{-1}$ in the experiment. 
This difference is attributed to the use of 
a minimal $sp^3$ basis in the calculation. 
However, our
calculations reproduce well the overbending of the LO 
band (i.e., the highest frequency is not
at the $\Gamma$ point, but at intermediate 
points between $\Gamma$M and $\Gamma$K). 
This has important consequences in the phonon
structure of the nanotubes, especially in the
zone-folding scheme, where the overbending leads to Raman 
active modes in the nanotubes which are higher in frequency than
those of graphene. Also, our results show good quantitative
agreement with experiment for the lower frequency band.
For example, for the out of plane transversal optical 
mode (labeled in the Fig.~\ref{dispersions} (a) as ZO), 
we obtain a frequency of 861 cm$^{-1}$, to be compared 
with the corresponding infrared actived mode in bulk graphite 
of 868 cm$^{-1}$. Acustic bands are also well reproduced.

The main features in the
phonon dispersion of Fig.~\ref{dispersions} (a) are
in good agreement with other density functional
calculations~\cite{Pavone,Kresse_grafito}, but differ from the phonon bands
obtained with empirical force constants models,~\cite{zf,Eklund}
especially around the M point.

Sound velocities are 
extracted from the slope of the acoustic branches. 
We obtain 24~km/s and 18~km/s for
the LA and the in plane TA branches, respectively.
These results agree very well with the sound velocities 
that can be extracted from the experimental 
data of Ref.~\onlinecite{Oshima} for the graphite 
(0001) surface phonons, $\approx$~24~km/s and 
$\approx$~14~km/s. 
The out of plane transversal band (ZA in the Fig.~\ref{dispersions} (a))  
has a zero sound velocity. 
Fitting the frequencies below 80~cm$^{-1}$ to a parabolic 
function $\omega=\delta q^2$, a value of
$\delta \approx 6~\times~10^{-7}~m^2s^{-1}$ is obtained, 
in very good agreement with previous estimations~\cite{Lucas}. 
\begin{figure}[!t]
\centering
\epsfxsize=0.9\linewidth
\epsffile{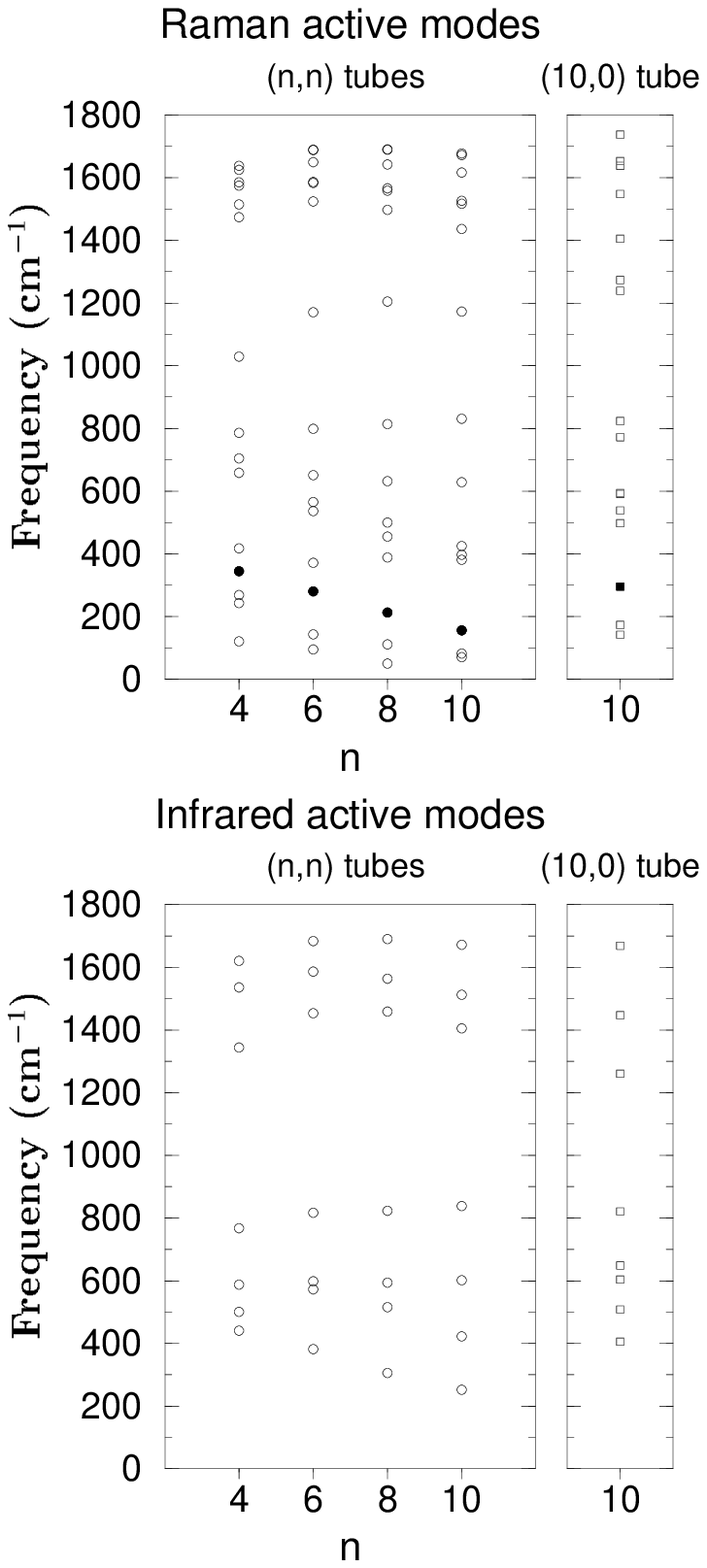}
\vspace{0.5 cm}
\caption[]{ {\it Ab initio} frequencies of the Raman and 
infrared active modes for a series of $(n,n)$ nanotubes, 
and for the  
$(10,0)$ nanotube. Filled symbols 
indicate the breathing mode.}
\label{Raman-Ir}
\end{figure}  
The sound velocities of the LA and TA branches allow us to calculate
back the in-plane stiffness (71.6 eV/atom) and the shear modulus
(40.3 eV/atom) of the graphene sheet, respectively. Both values 
are higher than the ones
obtained directly (see previous subsection) due to error propagation:
the elastic constants depend quadratically on the sound velocities, 
and these are delicate to obtain due to numerical
problems close to the $\Gamma$ point. The other way around, 
the sound velocity calculated using the experimental basal 
plane shear modulus of bulk graphite (c$_{66}$=0.44~TPa~\cite{Poison_grafito})
is 14~km/s, which is indeed quite close to our value.

From the quadratic behaviour of the ZA band 
we can also estimate the energy necessary to 
roll up the graphene plane to form the tubes.
It can be easily shown that the strain energy
per carbon atom can be approximated by 
$E_{st}=C/r^2$ where $C=\delta^2 m_c/2$, 
being $m_c$ the carbon 
atom mass and $r$ the tube radius.
The value obtained for C is 2.3~eV\AA$^2$/atom,
which is only a 7\% higher than the calculated 
constant for the (10,0) and (8,4) tubes, and 13\% higher
than the same constant for ($n,n$) tubes.

We next investigate the phonon structures of the
nanotubes considered here. Let us first comment on
the numerical errors in the calculated vibrational frequencies.
These are mainly originated from the residual forces in the
structural relaxation, the finite atomic displacements
in the force constant matrix calculation, and the finite grid
utilized in the integration of the Hamiltonian matrix
elements. We have estimated this numerical error from
the differences in the frequencies obtained
for the (10,0) tube by displacing two different
(but equivalent by symmetry) atoms for the calculation
of the force constant matrix. We find that the error
is about 10 cm$^{-1}$ for the high part of the spectrum
(frequencies higher than 1300 cm$^{-1}$), and could increase
up to 30 cm$^{-1}$ for some of the lower branches. 
The breathing mode and the acustic bands are more stable,  
showing errors of about 15 cm$^{-1}$. 
Uncertainties of the same magnitude have been 
reported by other authors performing similar 
calculations on graphitic 
systems.~\cite{Kresse_grafito,Miyamoto} 

Fig.~\ref{dispersions} (c) and (d)
show the calculated {\em ab~initio}
1D dispersion relations for the armchair (4,4) and (10,10)
tubes. The
zone-folding results for the tube (4,4), 
obtained from the graphene {\em
ab~initio} dispersion relations, are also displayed in the
panel (b) of the same figure.
The difference between the {\em ab~initio} and the zone-folding
frequencies are a consequence of curvature and
relaxation effects, and are therefore a measure of the
their importance in the phonon spectrum.

From the results of Fig.~\ref{dispersions} 
we see
that, 
apart from some small differences which we will 
analyze in the following, 
the general agreement between
the {\em ab~initio} results and the zone-folding
predictions is considerably good. This is the case
even for the tubes with smaller radii, where the
curvature effects in the phonon frequencies are
expected to be more important, and the zone-folding
scheme could start to break down.
The agreement is particularly good for the upper
part of the spectrum, and worsens for decreasing 
frequencies. This is mainly due to the failure of
the zone-folding approach to describe the breathing
modes and two of the acoustic bands of the tube
(those corresponding to motion in the directions
perpendicular to the tube axis).
In particular, within the zone-folding
scheme, the breathing mode appears with zero frequency,
and the two translational modes appear
with finite frequency. 
It should be noticed that these deficiencies
can be corrected~\cite{zf}, and
an analytical expresion can be obtained for the breathing
mode frequencies making similar assumptions as those
made in the zone-folding scheme: the use of
force constants from a graphene plane.  Jishi {\em et al.}\cite{zf}
showed that this model predicts a $1/r$ dependence of the 
frequency of this mode, regardless of chirality.

The deficiencies mentioned above are absent in the {\em ab~initio}
results, without having to resort to
additional corrections. In Fig.~\ref{dispersions} (c)
and (d)
we display with thicker
lines the phonon bands associated with the breathing and
twiston modes for the {\em ab~initio} results.
\begin{figure}[!t] 
\centering
\epsfxsize=0.9\linewidth
\epsffile{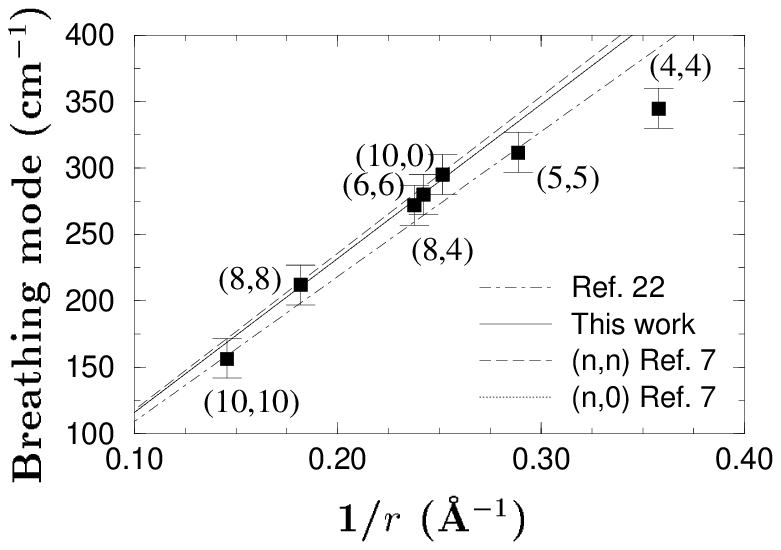}
\vspace{0.5 cm}
\caption[]{{\em Ab initio} breathing mode 
frequencies as a function of the inverse tube
radius, for (8,4), (10,0) and five $(n,n)$ tubes.
The continuous line is a linear fit to 
the data excluding the (4,4) and 
(5,5) tubes. Dot-dashed line shows the behaviour 
obtained by Jishi {\it et al.}~\cite{zf} using a 
force-constants model. 
Dashed and dotted lines show the result of the LDA calculations
by K\"urti {\it et al.}~\cite{Kresse_breathing} 
for the $(n,n)$ and $(n,0)$ 
tubes, respectively. In this scale the dotted line 
is hidden by the continuous one. 
Error bars as estimated (see the text).} 
\label{breathing}
\end{figure}

The twistons are torsional acoustic modes which 
have been proposed to be
of relevance for the peculiar linear dependence of
the of the electrical resistivity with temperature in the
metallic $(n,n)$ tubes.\cite{twistons}
These vibrational modes break the reflexion symmetry
of the tubes, and open a gap at the Fermi level,
producing a strong electron-phonon coupling, key to
the understanding of that behavior.
The sound velocity of these modes
is therefore important for the electronic properties
of the $(n,n)$ tubes. Our results indicate that the 
twiston mode sound velocity is lower than the 
corresponding value obtained for graphene 
(TA band) for all the studied tubes, and slowly diminish with
decreasing tube radius. For the (10,10) tube, the twiston 
sound velocity is 15~km/s, i.e. a 17\% lower than 
the value found in the graphene plane, and for the 
narrower tube (4,4) the value decreases to 13~km/s.

Fig.~\ref{Raman-Ir} shows the Raman
and infrared active modes at the $\Gamma$ point. These
have been assigned according to the $D_{nd}$ groups,\cite{zf}
which predict 7 infrared active modes and 15 Raman
active modes for the $(n,n)$ tubes, and 8 infrared and
16 Raman actives modes for the $(n,0)$ regardless
of their radius.
We note 
that, for the (10,0) tube, the highest Raman active mode in
the {\em ab~initio} calculation has a frequency larger than 
the correspoding frequencies for the $(n,n)$ tubes.
In our calculation this highest Raman active mode has a frequency
even larger 
than the predicted by the zone-folding scheme.
This is in contrast with the findings for the $(n,n)$ tubes.
In the (10,0) tube, this mode corresponts to an optical vibration in which
the displacement vector is parallel to the tube axis.
Although the position of this mode is quite sensitive
to the numerical precision of the calculation (and in particular
to the Brillouin zone sampling utilized), we can conclude that
this  tube presents optical frequencies
which are higher than those of $(n,n)$ tubes with comparable
radii.

\begin{figure}[!t] 
\centering
\epsfxsize=0.9\linewidth
\epsffile{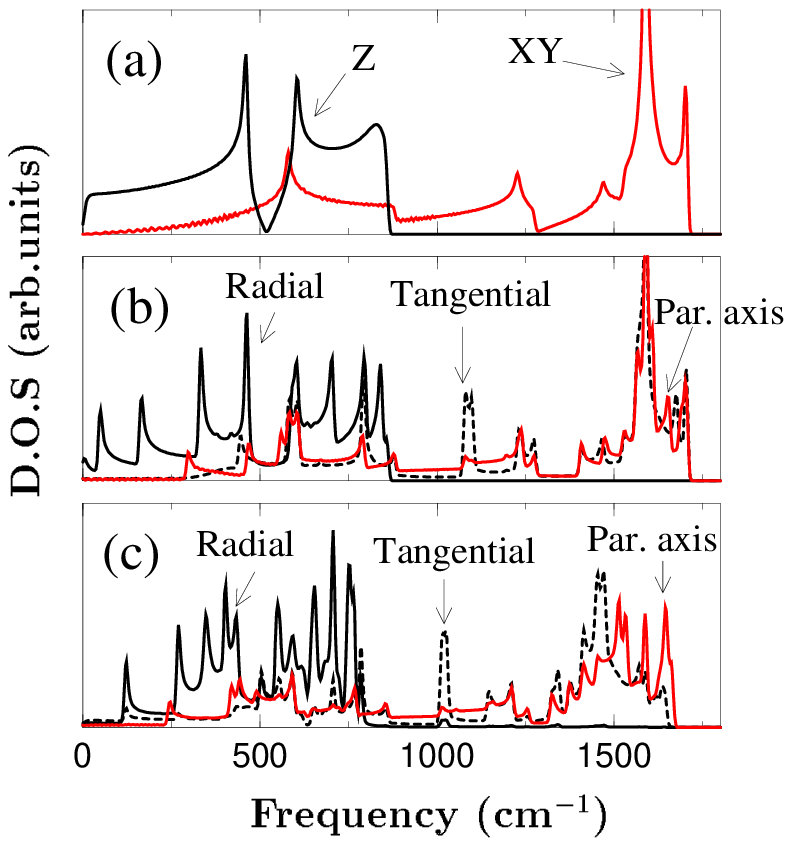}
\vspace{0.5 cm}
\caption[]{Vibrational density of states for (a) graphene, 
(b) the (4,4) tube in the zone folding approximation,
and (c) the {\em ab~initio} results for the same tube.
The curves are decomposed in the different directions
of the displacement vector. For graphene, $Z$ indicates
modes perpendicular to the plane, and $X$ and $Y$ within the
plane. For the (4,4) tube, the modes are decomposed in radial,
tangential and parallel to the tube axis.}
\label{dos}
\end{figure}

We have also analyzed the dependence of the breathing
mode frequency with the tube radius and chirality.
The results are shown in Fig.~\ref{breathing}.
The breathing mode is a $A_{1g}$ symmetry mode
(being therefore Raman active),
in which there is a monopolar inward and outward 
vibration of the atoms.
A simple approach, based on  the force constants derived
from the graphene plane, predicts a change of the
frequency of this mode as $A/r$, independently
of chirality, where $r$ is the tube radius and 
$A$=1092~cm$^{-1}$~\AA\ (Ref.~\onlinecite{zf}).
Recent LDA frozen phonon calculations~\cite{Kresse_breathing}
of this mode
confirm the prediction of the $r^{-1}$ behaviour, 
the constant $A$ having a weak dependence on chirality,
1180~cm$^{-1}$~\AA\ and 1160~cm$^{-1}$~\AA\ for $(n,n)$ 
and $(n,0)$, respectively. Our calculations confirm the previous results,
indicating once again that the effect of
curvature on the value of the force constants
is small, even for the small radii tubes considered
here. Only the (4,4) tube presents an important 
deviation from the predicted behaviour, with an
appreciable decrease in the breathing mode frequency.
This effect is already noticeable in the (5,5) tube,
although to a smaller extent. This effect can be understood
as a consequence of the hibridization changes and the
decrease of the $\pi$ interaction, induced by the
curvature. A $r^{-1}$ fit of the results for tubes
with radius greater than 3.8~\AA\ gives a value of
1160~cm$^{-1}$~\AA\ for the constant $A$, in very 
good agreement with Ref.~\onlinecite{Kresse_breathing},
as clearly shown in Fig.~\ref{breathing}.
The possible chirality dependence of the breathing mode, 
if any, is well below the resolution of our data.
As pointed out in Ref.~\onlinecite{Kresse_breathing}, 
the value of $A$ can be estimated from 
the stretching constant of the graphene plane
neglecting all the possible effects of curvature.
Taking our calculated value of 60~eV/atom (see previous sections)
for this elastic constant, we obtain a value $A$=1166~cm$^{-1}$~\AA.

\begin{figure}[!t] 
\centering
\epsfxsize=0.9\linewidth 
\epsffile{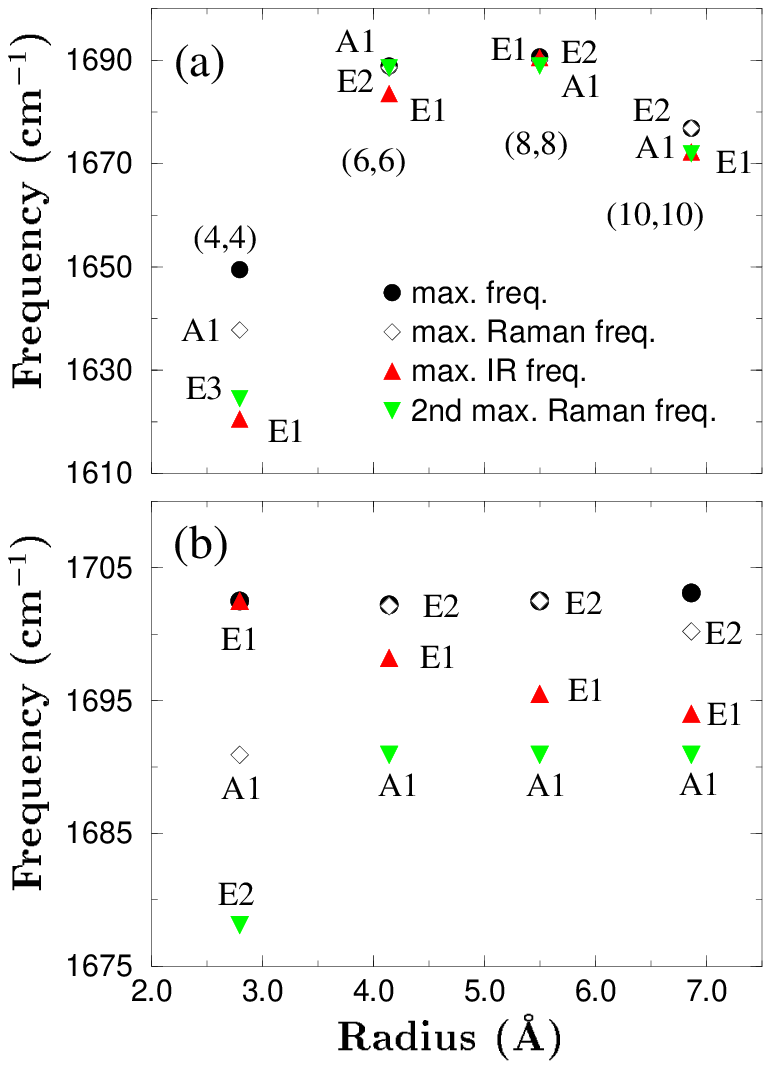}
\vspace{0.5 cm}
\caption[]{Frequencies for the higher optical modes for different
tubes, (a) {\it ab initio}, 
and (b) from the zone folding scheme. We show the frequencies
of the highest vibrational mode, the two highest
Raman active modes and the highest Ir active mode.
The symmetry of each mode is also shown.}
\label{maxw}
\end{figure}

One of the most important differences between the
{\em ab~initio} and zone-folding frequencies
is a general softening of frequencies when curvature
effects are taken into account. This is specially
clear for the higher frequency bands, but is also
observable for the intermediate frequencies.
The shift is not uniform, and therefore the distribution
and, in some cases, the ordering of the Raman
and Ir active modes are affected by curvature effects. 
This effect is most evident in the (4,4) tube, where the 
curvature is more pronounced.
We show in Fig.~\ref{dos} the vibrational density of
states for the (4,4) tube, comparing the {\em ab~initio}
and the zone-folding results.
The figure shows the decomposition of the modes in 
radial and tangential (parallel and perpendicular to
the tube axis). We see that the radial modes
correspond to frequencies below 800 cm$^{-1}$, 
corresponding roughly to the out of plane  
bands of graphene. The softening of the frequencies
in the {\em ab~initio} calculation are apparent in
this figure, specially for the higher frequency modes.
Also the upper limit of radial vibrations is lowered
by about 100 cm$^{-1}$ in the {\em ab~initio} calculation
compared with the zone-folding results, for the (4,4) tube.

Several works~\cite{Rao,Kasuya} have recently made use of
the frequency of the higher optically active modes 
as an experimental signature of the tube radii.
In graphite there is only one Raman active optical mode
(with zero wave vector) at 1580 cm$^{-1}$, which in our calculation
appears at 1690 cm$^{-1}$. In the nanotubes, 
it splits into multiple peaks which are originated in the 
zone folding of the graphene bands. The frequencies of these 
$q_z$=0 tube modes depend on radius and chirality.
In the zone-folding scheme, these modes sample the corresponding
optical bands of graphite, with wave vectors $q_n=n/r$ ($n=0,1,2,...$)
along the circumference direction.
Of these, only a small number (independent of the tube
radius) are Raman or Ir active.\cite{zf,Eklund}
Kasuya {\it et al.}\cite{Kasuya} were able to measure the frequency
of the highest Raman modes for tubules with different radii,
and found that these corresponded to the $n=1$ LO band,
and $n=1$ and $n=2$ TO bands of graphite (in order of
decreasing frequencies). They found the value of the observed
frequencies to be in excellent agreement with the
direct zone-folding results. The radius of the narrower tubes
in their sample was  5.5 \AA, which, assuming an
armchair conformation, would correspond to the (8,8)
tubule. It is, therefore, interesting to see whether
these results are modified for tubes with smaller radii,
as was the case for the breathing mode discussed above.

Fig.~\ref{maxw} (a) shows the frequencies of the
higher optical modes obtained from the {\it ab-initio}
calculation, whereas Fig.~\ref{maxw} (b) shows the results from
the zone-folding approach. Several facts are worth noticing in the 
comparison. As was the case for the breathing mode, the
(4,4) tube deviates very significantly from the
zone-folding behavior. The maximum {\it ab-initio} frequency 
is about 50 cm$^{-1}$ lower than the zone-folding prediction, and 
the symmetries and activities of the higher frequency peaks
are very different.
For the larger tubes, it seems that the {\it ab-initio}
results tend to confirm the symmetry assignments of the
zone-folding approach, supporting the analysis 
of experimental results in terms of the simple 
zone-folding scheme. 
The apparent softening of the maximum frequencies for 
the (10,10) tubule in Fig.~\ref{maxw} (a) is due to numerical
error (the 10 cm$^{-1}$ error bar discussed above, possibly slightly
larger for the largest tube). The maximum frequency for this tube 
should approach the one obtained by the zone-folding method since the 
curvature is the smallest. This serves as a measure of the accuracy
of our calculation.

\section{CONCLUSIONS}

We have presented the results of {\it ab~initio} calculations for 
single-wall carbon nanotubes with different chiralities and radii, 
addressing structural, elastic, and vibrational properties.
The cases studied include $(n,n)$ tubes (with $n$ ranging from 4 to 10), 
and  the (8,4), and (10,0) tubes. We have also presented a detailed 
comparison with the results of other usual 
approaches like elasticity theory for elastic properties 
or the zone-folding approach for the vibrations.
Our results serve to validate most of the predictions
of these simpler theories, and to point out their limits of aplicability.
The following conclusions can be drawn from our results:

$(i)$ Relaxation effects due to the tube curvature are small in general.
The inequivalent bonds in a tube enhance their differences in
bond length and angles with decreasing tube radius. The symmetry
inequivalent bonds give rise to a small shift in the Fermi surface 
location for the ($n,n$) tubes, mainly related to the lower symmetry
of the tubes as compared with graphene, and only slightly modified
by the structural relaxation.

$(ii)$ The strain energy follows the $\alpha \over r^2$ 
law expected from elasticity theory
quite accurately for tubes as narrow as (4,4).  
For armchair tubes, which have slightly lower strain energy 
than other chiralities, the
constant has a value of $\alpha=2.00$~eV\AA/atom. 

$(iii)$ Sensible definitions of the Young modulus are used for 
two different geometries: multi-wall and single-wall tubes. 
In the former case the values
are very similar to the one of graphite.
Single-wall tubes show values smaller than graphite.
In any case, we propose the elastic constant per unit mass 
as the relevant quantity, since it does not depend on
the geometry of the system. This is shown to be quite 
similar to the correspondent quantity in graphite,
for all the studied tubes,
and larger than in any known fiber.

$(iv)$ The Poisson ratio also retains
graphitic values except for a possible slight reduction for small radii.
It shows a chirality dependence: ($n,n$) tubes display smaller 
values than (10,0) and (8,4). Our results for the $(n,n)$ tubes
are consistent with the experimental basal Poisson ratio of
graphite.

$(v)$ The phonon bands behave as expected from simple
schemes, except for slight deviations which become more important for the 
narrower tubes. The zone-folding analysis gives a 
good qualitative (and sometimes quantitative) picture of
many of the properties studied here, except for known defficiencies
in the low-frequency vibrational spectra. For the smallest 
radii, the zone-folding description
of the high-frequency vibrations is insufficient, too.

$(vi)$ The breathing mode follows the $A/r$ law
predicted by graphene-derived force-constants calculations. 
The obtained value of $A$ is consitent with that calculated
from the in plane stretching elastic constant of graphene. 
It, however, seems to soften with respect to the
expectations for the smallest radii tubes, like (4,4).
A similar softening is observed for the twiston modes, whose
sound velocity diminishes for decreasing radii.

$(vii)$ The high-frequency optic modes are sensitive to the kind of tube
and to its radius. The frequencies of the highest modes tend to diminish
with decreasing radii by effect of the curvature.

{\em Acknowledgments: } 
We are grateful to Javier Junquera for giving us the possibility of
performing k-sampling tests with his code.
We acknowledge financial support from DGICYT under grant PB95-0202,
and European Community TMR contract ERBFMRX-CT96-0067 (DG12-MIHT).


\begin{references}
\bibitem{distancias}
We have found that the difference between the 
inequivalent bond-lengths is extremely sensitive to 
the k-point sampling. The behaviour shown in the 
Fig.~1 (a) has been checked in two ways: (i) 
by comparing
the tube results with the relaxations of a graphene 
sheet with an identical k-point sampling, and (ii) 
using more complete k samplings in the tube calculations. 
In both case the 
same behaviour as presented in Fig.~1 (a)
was found when the bond-lengths are expressed
in units of the corresponding bond distances for graphene 
calculated with equivalent sampling. The absolute lengths are
more difficult to converge, being inverted for the samplings used
in Ref.~\onlinecite{Kresse_breathing}.
\bibitem{pw_pois} This data is in agreement 
with ab~initio pseudopotential
plane-wave calculations of the poisson ratio in carbon nanotubes.
The computed values of 0.151 and 0.180 for the 
(5,5) and (10,0) nanotubes are 
in excellent agreement with the present calculations of 
0.14 and 0.19, respectively.
\end{references}
\end{document}